Imaging the Inner Structure of a Nuclear Reactor by Cosmic Muon Radiography

January 11, 2019

Hirofumi Fujii[1], Kazuhiko Hara[2], Shogo Hashimoto[2#], Kohei Hayashi[1], Fumiaki Ito[2], Hidekazu Kakuno[3], Hideyo Kodama[1], Kanetada Nagamine[1], Kazuyuki Sato[2*], Kotaro Satoh[1], Shin-Hong Kim[2], Atsuto Suzuki[1$], Takayuki Sumiyoshi[3], Kazuki Takahashi[2#], Yu Takahashi[2&], Fumihiko Takasaki[1], Shuji Tanaka[1] and Satoru Yamashita[4]

[1] *High Energy Accelerator Research Organization (KEK), Tsukuba, Ibaraki 305-0801, Japan*
[2] *University of Tsukuba, Tsukuba. Ibaraki 305-8571, Japan*
[3] *Tokyo Metropolitan University, Japan*
[4] *University of Tokyo, Japan*

[*]Present address: Iwate Prefectural University
[†]Present address: Toshiba Co., Ltd.
[$]Present address: JAEA
[&]Present address : JAXA
[#]Present address : NEDO

**Abstract**

We studied the inner structure of the nuclear reactor of the Japan Atomic Power Company (JAPC) at Tokai, Japan, by the muon radiography. In this study, muon detectors were placed outside of the reactor building. By detecting cosmic muons penetrating through the wall of the reactor building, we could successfully identify the objects such as the containment vessel, pressure vessel, and other structures of the reactor. We also observed a concentration of heavy material which can be attributed to the nuclear fuel assemblies stored in the nuclear fuel storage pool.

Subject Index [Insert subject index codes here]

1. **Introduction**

The nuclear power reactors at the Fukushima Daiichi, Japan, were seriously damaged by the gigantic earthquake followed by Tsunami in March, 2011. Decommissioning of the damaged reactors has to be made and establishing the procedure of decommissioning is one of the urgent tasks. As the reactor buildings and their surrounding area of the Fukushima-Daiichi are highly radioactive, close inspection of reactors is very much limited. It was then proposed to use cosmic muons to diagnose the reactor status from outside of the reactor building by Cosmic Muon Radiography [1], [2]. It is to measure millions of cosmic muons which penetrate the reactor and to figure out detailed structure of the reactor. The cosmic muon detectors are placed outside of the damaged reactor building. From the observed muon distribution, it was hoped to visualize the inner structure of the reactors.

In order to demonstrate the usefulness of the cosmic muon radiography technology in visualizing the inner structure of the reactor, we made a measurement of cosmic muons penetrating through the nuclear reactor of the Japan Atomic Power Company (JAPC), Tokai, Ibaraki, Japan. A picture of the investigated JAPC reactor building is shown in Fig. 1. The main building is about 40 m high and 40 m wide. This reactor is a GM MKII type and its conceptual structure image is given in Fig. 2. The measurement was made from February 2012 to December 2013. During this period, the reactor ceased generating electricity and nuclear fuels were taken out and stored in the storage water pool (NFSP). We reported results of our first measurement in [3], concluding that (i) no fuel is loaded in the reactor and (ii) a cluster of heavy object is identified in the location of NFSP. Since February 2013, we added the second system at a horizontal view angle nearly 90° with respect to the first system. This report describes the impact of this addition: we succeeded in 3D imaging of the cluster of fuel objects placed in the NFSP.

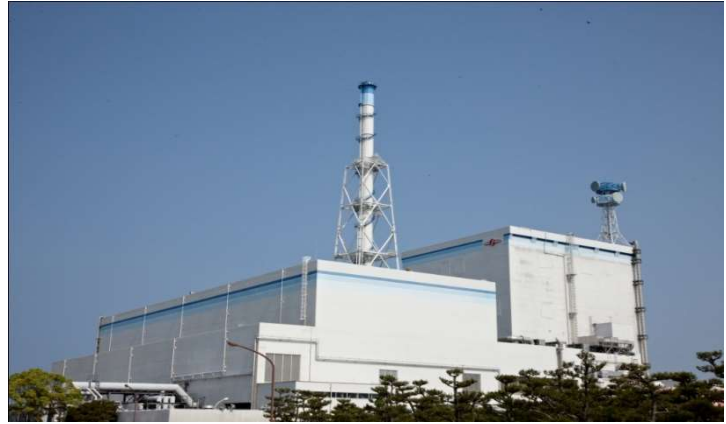

Fig.1. The investigated nuclear power reactor of the JAPC at Tokai (courtesy of JAPC)

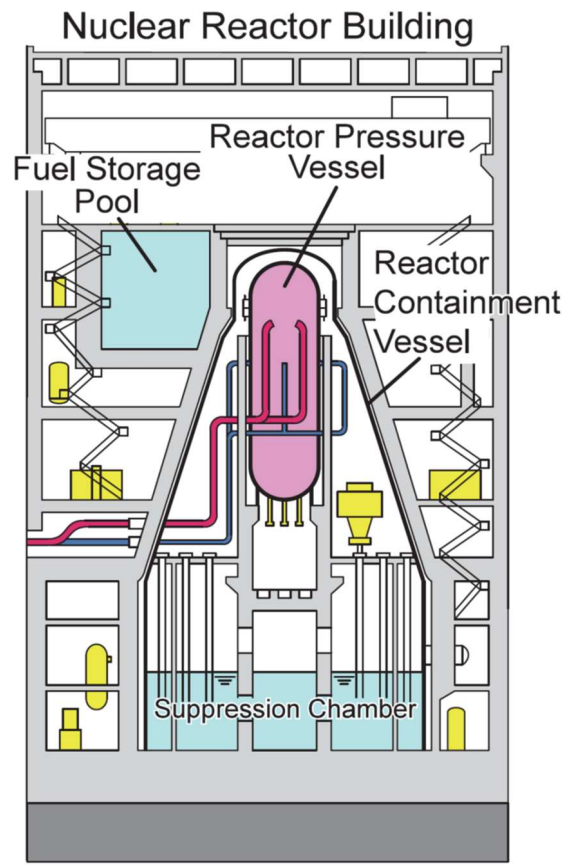

Fig. 2. Conceptual structure image of the Reactor of the JAPC at Tokai

## 2. The Muon Detection System

Since the details of the muon detection system are given in [3], we describe briefly the detection system supplementing the previous description. The major difference is the addition of the second muon telescope, made of two X-Y units of the plastic scintillation counters (PSC). The two X-Y units are separated by 1.5 m. We call this telescope as MT2. The MT2 is not equipped with a magnet unlike MT1 [3] of the first muon telescope system.

The basic unit of the PSC is composed of a scintillator bar of 1 m length and 1×1 cm$^2$ in cross section with extruded hole at the center [4], where 1 mm$\phi$ wavelength-shifter fiber [5] is embedded. One of the fiber end is coupled with a multi-pixel photon counter (MPPC) [6]. The efficiency of the PSC was evaluated [7] using cosmic muons by a test detector system fabricated using the same PSCs. The test detector system has a five-layer structure with 10 PSCs assembled in each layer as shown in Fig. 3. We evaluated the detection efficiency of a particular PSC in the middle layer $D_T$ while the traversing muon track was identified by the PSCs painted blue in the layers of $D_1$, $D_2$, $D_3$ and $D_4$.

We counted the number of events, N1, with the condition of having signals from the PSCs painted in blue in coincidence, and the number of events, N2, by requiring an additional signal from the PSC in the layer of $D_T$ in coincidence. The detection efficiency for cosmic muon was estimated from the ratio, N2/N1, by changing the threshold voltage to the $D_T$. As one sees in Fig. 4, the detection efficiency of the PSC approaches to unity for small value of the threshold $V_{th}$ and drops to almost zero for $V_{th}$ higher than 1 V. Similar study was made for the case where the muon traverses the gap region of the plane (see Fig. 3). Taking into account of the gaps between the PSCs, the overall detection efficiency for the muons of normal entry is estimated to be better than 95%. The efficiency is expected to be better for the muons traversing with angles.

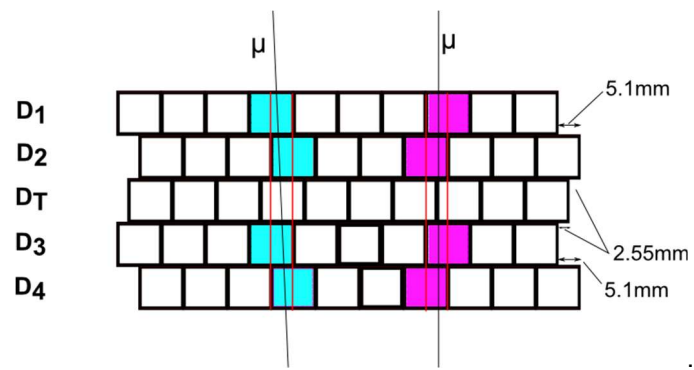

Fig. 3: Cross sectional view of the PSC arrangement of the test detector system. The PSCs in the layer marked as $D_T$ are used to estimate the detection efficiency. Two cases are shown: Case 1 (blue) muons traversing through the center of a PSC in the layer of $D_T$ and Case 2 (purple) muons traversing through the boundary region of adjacent PSCs.

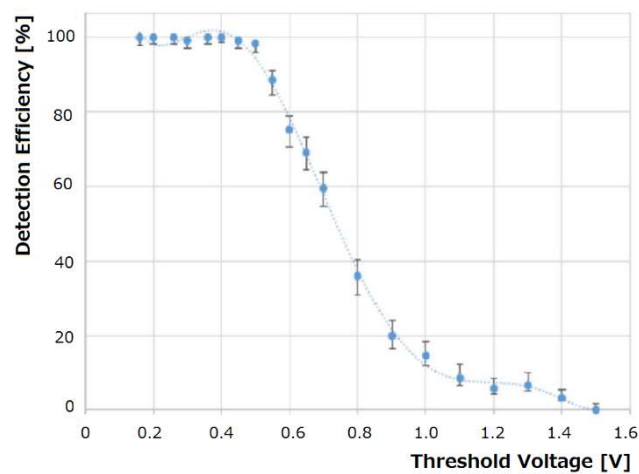

Fig. 4: Detection efficiency of the muon detector for the cosmic muons as a function of the threshold voltage (in volt) applied to the MPPC signal.

### 3. Imaging the reactor of the JAPC, Tokai

*3.1. Location of the muon telescopes*

The previous results in [3] were obtained with one muon telescope, MT1, placed 63 m away from the reactor center at the location MT1-1 shown in Fig. 5.

We continued the observation by changing slightly the location of MT1 to MT1-2 and by deploying another telescope, MT2. The locations were chosen in consultation with the JAPC. The telescope MT1 was placed on the ground outside of the reactor building 61-63 m away from the reactor center. MT1-1 was directed to the center of the pressure vessel, where the nuclear fuel is loaded when the reactor is operational. When it is not operational, the fuel is usually moved to the NFSP adjacent to the reactor core. The MT2 was placed on the roof of a building adjacent to the reactor building, about 30 m away from the reactor center with its elevation about 10 m high from the ground level.

The data taking in this detector configuration started in February 2013. Combining the images obtained by the MT1 at the places, MT1-1 and MT1-2, and MT2, we reconstructed three dimensional image of the nuclear fuel assemblies stored in the NFSP. The numbers of events used for the analysis are about 15 million at MT1-1, 12 million at MT1-2, and 8 million at MT2.

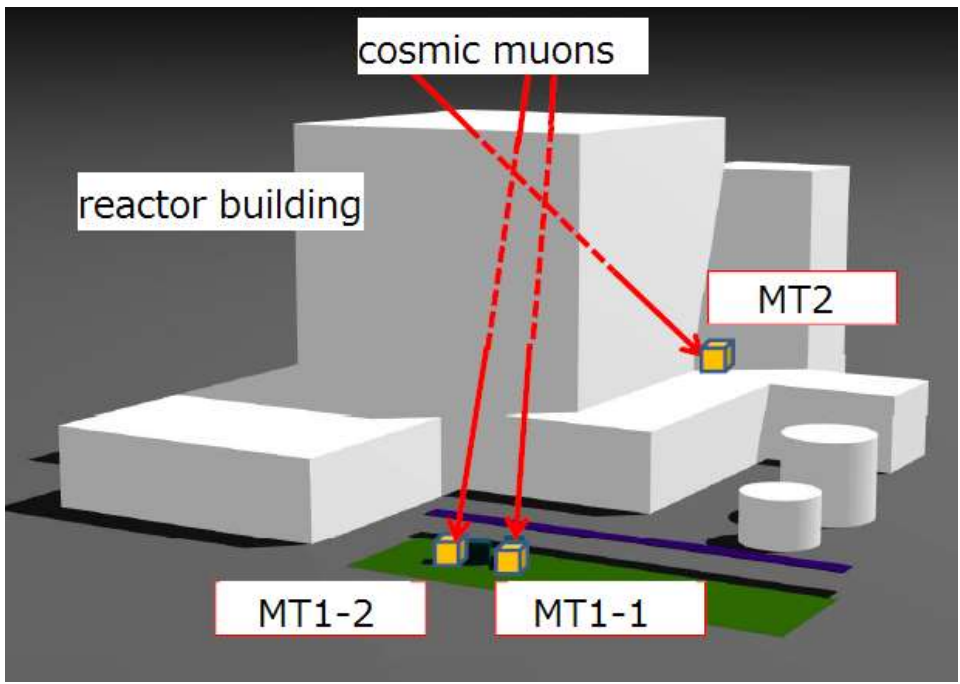

Fig. 5: Locations the Muon Telescopes (marked as the orange cubes) deployed at the nuclear power reactor of the JAPC.

*3.2. 2-Dimentional Images of the reactor*

From the cosmic muon distributions observed at the three detector positions, we reconstructed the inner structure of the reactor. The reconstruction of the reactor structure was made in the following steps:

1) "Detector Direction" is defined as the direction connecting the centers of the two XY-units in an MT. As we have four XY-units in the MT1, two of them are in front of the magnet and other two are befind [3], we define the Detector Direction using the front X-Y unit pair only.
2) Define an "Image Plane" which is perpendicular to the Detector Direction at the position of the XY-unit in rear.

3) Draw the line of incoming cosmic rays from the hit positions recorded by the XY unit pair. Then, we calculate the direction tan(θx) (tan(θy)) of the line determined by the horizontal (vertical) coordinate difference of the two hit positions divided by the spacing of the two. (Notice: we allow only one cluster hit in each XY-unit and threfore only one single line can be reconstructed.) We then calculate the position where this line traverses the Image Plane. By going through this procedure for all observed muons, we can depict an "Image of the Reactor" projected on the Image Plane.
4) We obtain the Image of the Reactor each for the data obatined at MT1-1, MT1-2, and MT2.

Figure 6 shows the image of the reactor obtained by the MT1 at the position MT1-1. The acceptance-corrected event rate is shown in a degree of darkness. The darkenss scale, indicated in the figure, is defined as the negative of the logalithm of the ratio R of the number of observed muons to that expected for given zenith angle θ and azimuthal angle ϕ,

$$-\ln R = -\ln [N_{obs}(\theta, \phi)/N_{exp}(\theta, \phi)], \qquad (1)$$

where the expected number of muons $N_{exp}(\theta,\phi)$ is the number of cosmic muons when there are no obstructing materials in the paths of incoming muons. More details are given in Sec. 3.3. The darker the spot is, the lesser the event rate is. The azimuthal angle at the center of the image is about 25° while the detector's azimuthal angular coverage is from 0° to 40°.

The pressure vessel, containment vessel, some timbers and girder beams of the reactor building and the wall of the building are marked by red points in the figure. Note that the horizontal girder beams are quite tilted in the image. One notices that there is no heavy object corresponding to the nuclear fuel at the loading zone in the pressure vessel. The dark region to the left of the containment vessel can be attributed as the NFSP.

Figure 7 shows the similar image obtained by the MT1 at poistion MT1-2.

Figure 8 is the similar image obtained by the MT2. The NFSP is seen behind the containment vessel.

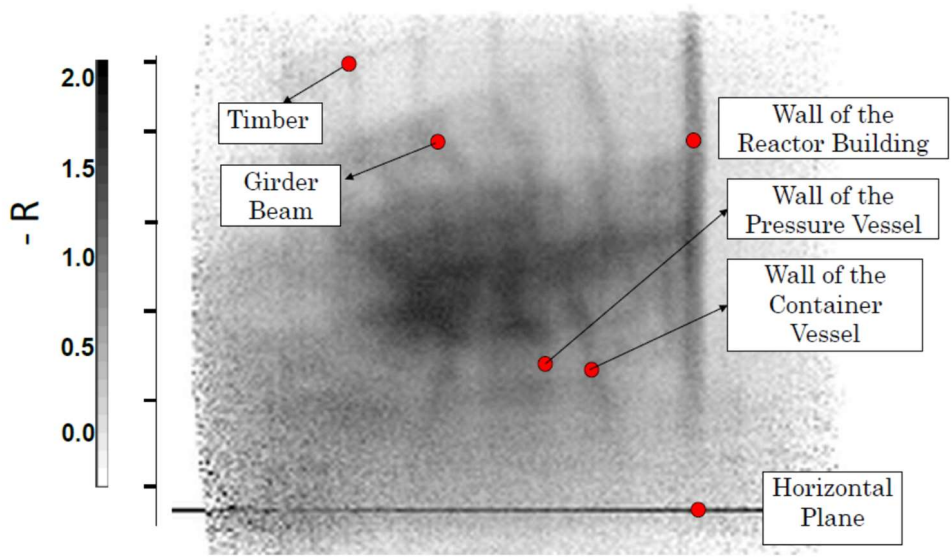

Fig. 6: Image of the reactor complex obtained from 15 million muons observed by the MT1-1. The darkness, corresponding to the magnitude of the attenuation of the muon flux, is defined in the text.

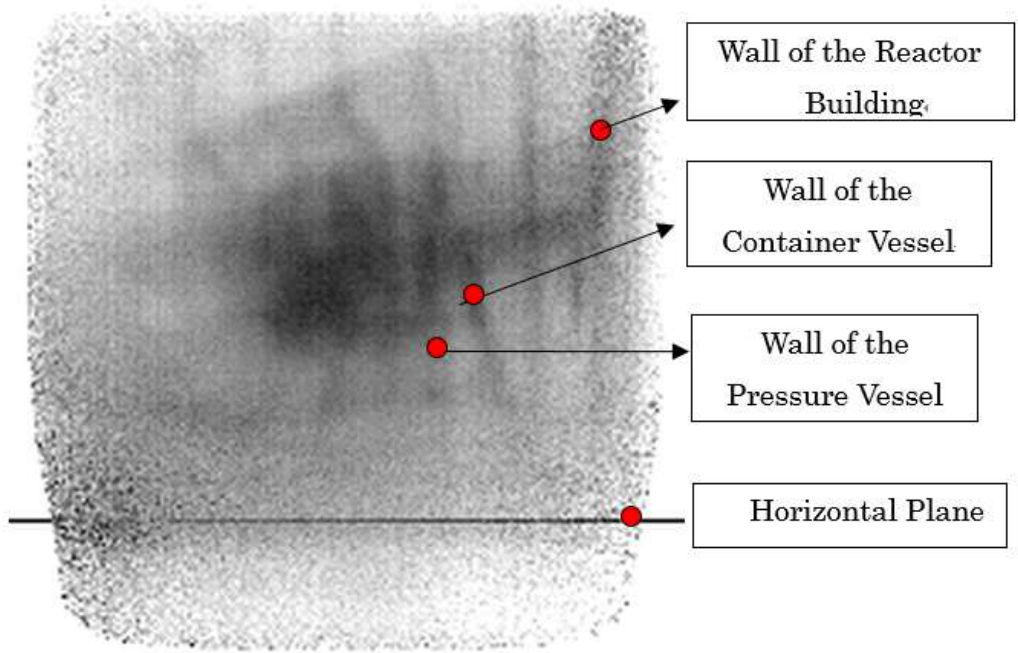

Fig. 7: Image of the reactor complex viewed by the MT1 at the position MT1-2  The darkness of the image is given in the same unit as in Fig. 6.

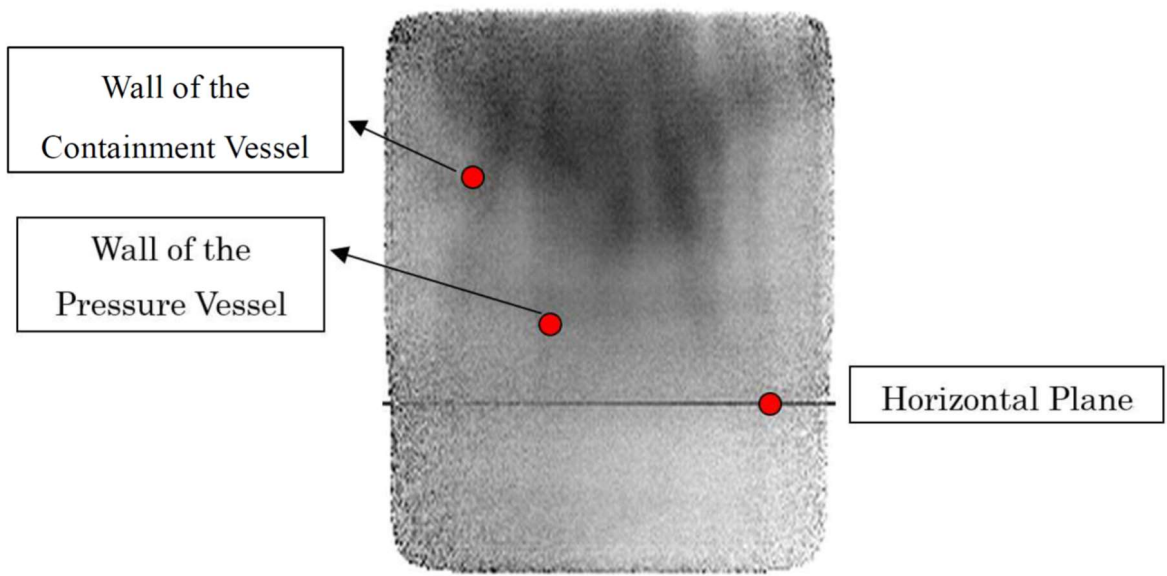

Fig. 8: Image of the reactor complex viewed by the MT2. The darkness of the image is given in the same unit as in Fig. 6.

Figure 9 shows the angular acceptance of the MT1-1. The MT-1 covers the region from -10° to +45° in θ and -30° to +30° in ϕ. The image gets blured at shallow θ angles partly because the cosmic muon flux decreases, especially in low momentum component which is essential in the muon radiography. Figure

10 is the angular coverage viewed by the MT2. It has a similar angular coverage as the MT-1 with slightly smaller coverage in θ.

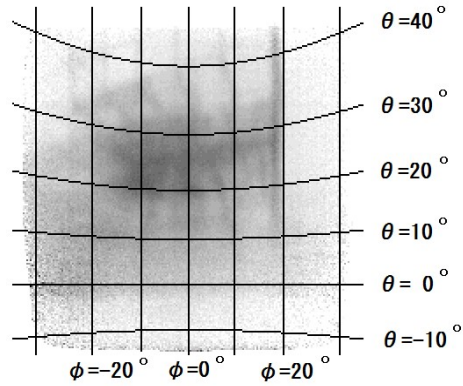

Fig. 9: Angular coverage of the image of the reactor viewed by the MT1-1.

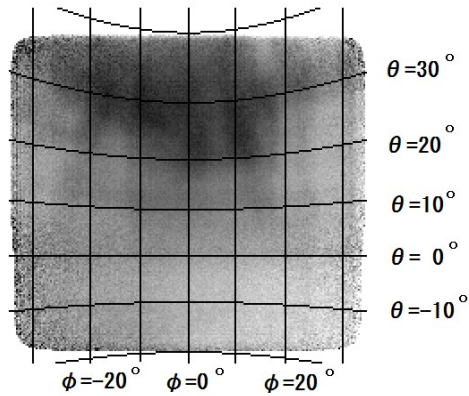

Fig. 10: Angular coverage of the image of the reactor viewed by the MT2.

*3.3. 3-D Image of Heavy Object in the NFSP*

In order to obtain a three-dimentional image combining the three 2-D images, we normalize the observed numbers of muons $N_{obs}(\theta,\phi)$ for given angles of θ and φ by dividing them by the expected numbers of muons.

The expected numbers of muons, $N_{exp}(\theta,\phi)$, are given by the following equation:

$$N_{expect}(\theta, \phi) = F(\theta)\, \Delta S\, \Delta\Omega\, T, \qquad (2)$$

where $F(\theta)$ is the comic ray flux given as a function of the zenith angle as

$F(\theta) = A[(1-r)\cos^2\theta + r]$
with A= 60 and r=0.025.

$\Delta S$ is the effective detection area, $\Delta\Omega$ is the solid angle for detecting muons and $T$ is the data taking time. $F(\theta)$ is given in unit of 1/str/m²/sec. Here we assumed that there is no azimuthal angle dependence of the cosmic ray flux such as the east-west dependence. This formula of the cosmic ray flux is verified with the cosmic ray flux measured with the MT1 at KEK where the MT1 was placed on the ground with no heavy objects surrounding it. Figure 11 shows the measured cosmic muon flux in black dots as a function of the zenith angle $\theta$, compared with the fomula, $F(\theta)$, shown in blue solid curve. One sees that the calculated flux, $N_{exp}(\theta, \phi)$, reproduces a general feature of the observed data, although there is a certain discepancy. This discrepancy, however, is in the large $\theta$ range and does not distort the imaging the NFSP region. We note that this formula is consitent with the description given in the Particle Data Book [8].

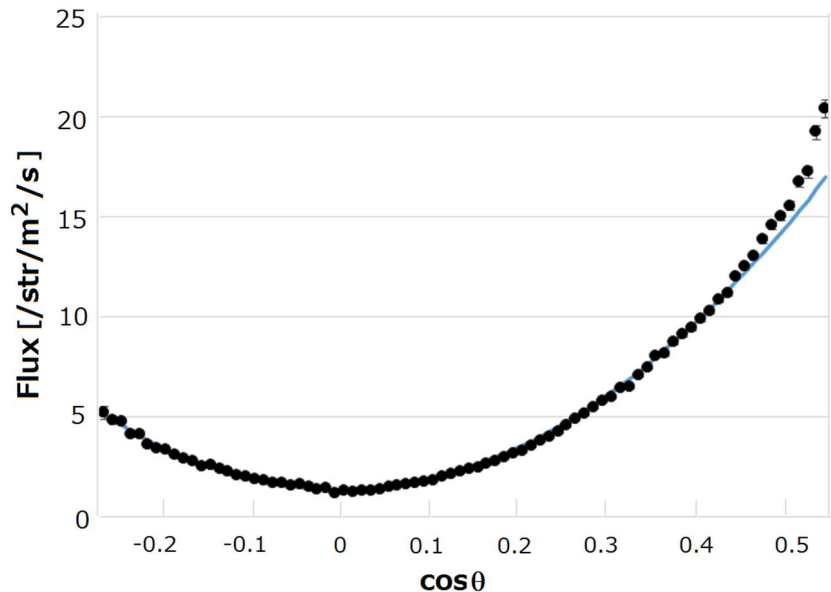

Fig. 11: Observed cosmic ray flux as a function of azimuthal angle with no obstacle material along the muon path. The solid blue curve is a calculation by the formula, $F(\theta)$.

As we have confirmed [6] that the present muon radiography system can image the inner structure of the reactor building from outside of the building, our next step is to locate the nuclear fuel materials stored in the NFSP and obatain its shape three dimentionally.

The starting point is to look for spots of large suppression in the muon flux in the 2-D images.To search for the area of significant suppression, we calculate the ratio of the numbers of observed comic muons to those expected $R= N_{obs}(\theta, \phi)/N_{exp}(\theta, \phi)$. Then, we mark those points as "highly suppressed" when ln(R) values are smaller than the threshold values. The thresholds are set for the three 2-D images by MT1-1, MT1-2 and MT2 at -1.4, -1.5 and -1.8, respectively. We then mark those "highly suppressed" points as shown in Figs. 12, 13 and 14 in blue points.

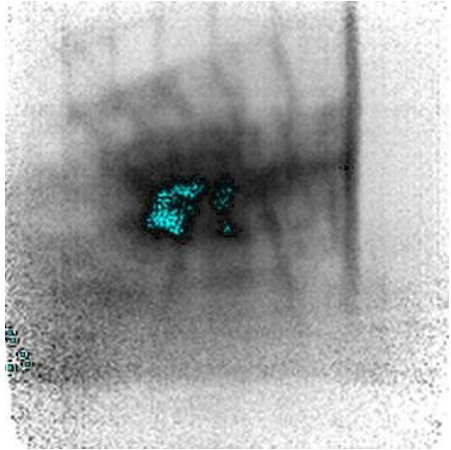

Fig. 12. Spots of significant cosmic ray suppression, $\ln(R) < -1.4$, marked in blue superimposed on the image of Fig. 6.

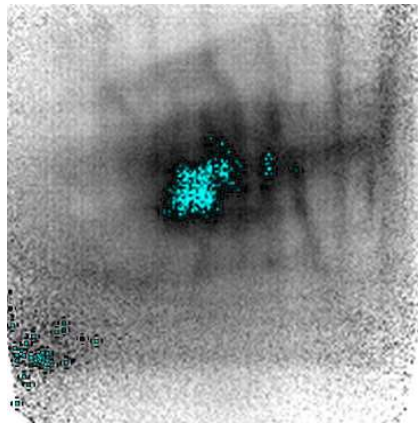

Fig. 13. Spots of significant cosmic ray suppression, $\ln(R) < -1.5$, marked in blue superimposed on the image of Fig. 7.

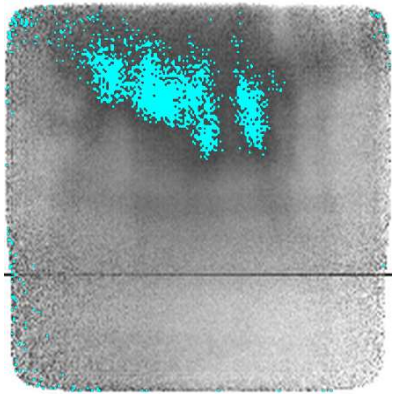

Fig. 14. Spots of significant cosmic ray suppression, $\ln(R) < -1.8$, marked in blue superimposed on the image of Fig. 8.

We observe a concentration of "highly suppressed spots" at the upper left side of the containment vessel in Figs. 12 and 13, and at the upper part of the containment vessel in Fig. 14, where the NFSP is located.

In order to obtain a 3-D image of the "highly suppressed spots", the NFSP volume is divided into blocks of 1 m cube, and we assign the blocks to be heavy if two or three out of the three 2-D images of "highly suppressed spots" point to the identical block. The reconstructed shape is displayed in green color in Fig. 15 together with the water in the NFSP in blue, viewed at 20 different angles around the center axis of the containment vessel. We observe two clusters of heavy blocks in the NFSP. We show one of the 20 images in a magnified scale in Fig. 16. It could be speculated that one cluster is of the spent nuclear fuels and the other of the fuel assemblies to be used for power generation. We successfully located and obtained the 3-D image of the fuel.

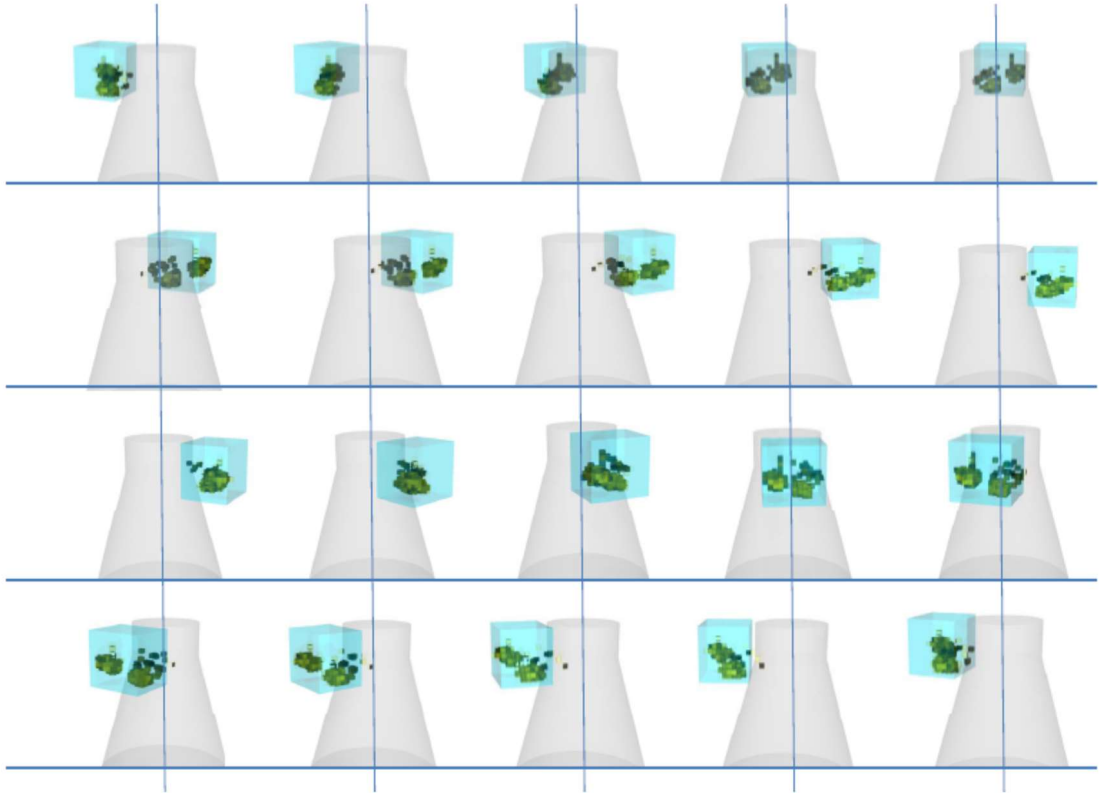

Fig. 15. Reconstructed 3D images of the spots of significant suppression of cosmic muon flux. The images are shown at different view angles varied at a step of 18°. The gray cone-shape is the containment vessel and the blue cubic area is the NFSP.

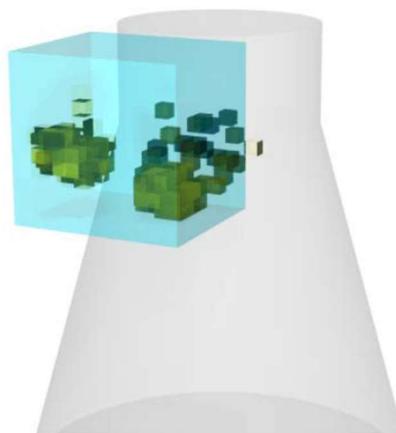

Fig. 16. One of the images of "spots of high absorption".

*3.4. Evaluation of Mass of the Heavy Object in the NFSP*

Evaluation of the mass of the heavy object is the ultimate subject of the muon radiography. Since we identified the shape of the heavy cluster, the average density is the quantity to be derived. The property that the muon absorption in material is dependent on the density times the path length of muons (density-length) is used to derive the average density. However, the absorption is also dependent on the muon momentum, and the muon energy spectrum becomes harder at lower zenith angles [8], which needs to be taken into account.

The zenith angle dependence of the muon absorption to density-length relation was derived using the measured muon transmittance data through the containment vessel, the pressure vessel and other reactor building materials but excluding the area of the NFSP. The density-length along such objects was calculated from the known materials and dimensions. Figure 17 shows the calculated density-length distribution viewed by MT2.

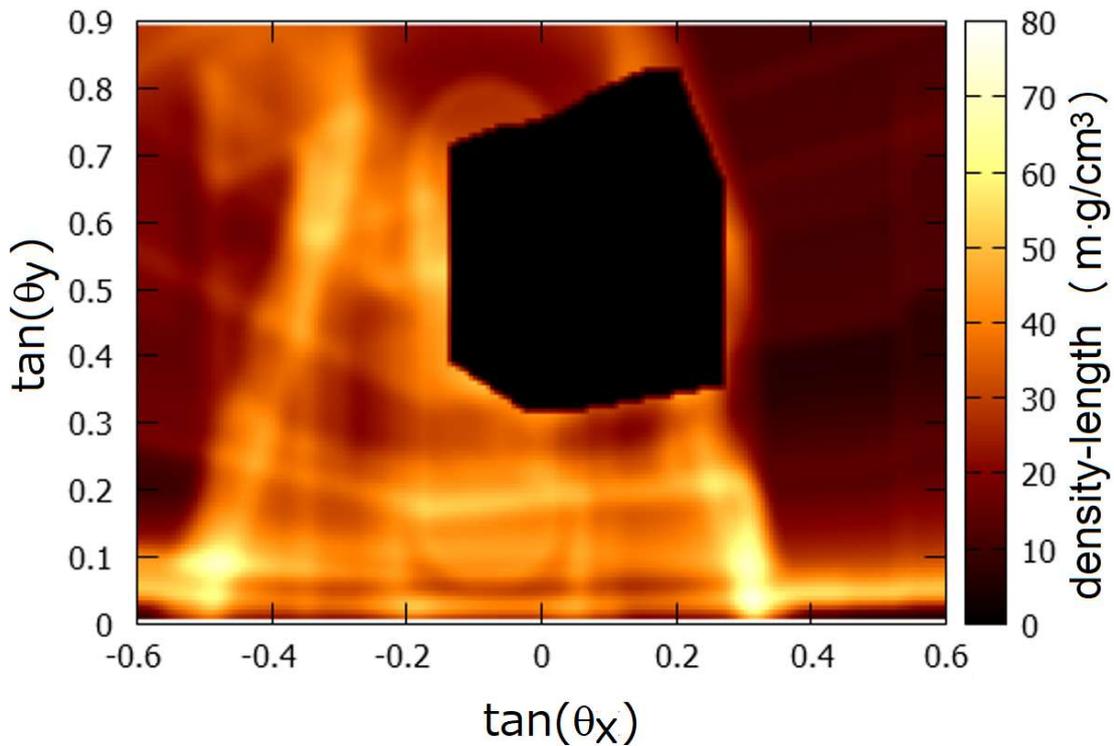

Fig. 17. Density-length distribution seen by the MT2 calculated using known density and dimension values. The coordinates are the horizontal and vertical directions (see Sec. 3.2) of the MT2. The area corresponding to the NFSP (black) is excluded in this calculation.

The transmittance of the muon, $T_m$, is found to be approximated well by the following equation,

$$-\ln(T_m) = a(\theta)\sqrt{L_D}, \qquad (3)$$

where $L_D$ is the density-length, and $a(\theta)$ is the zenith angle dependent coefficient. Figure 18 shows an example of relation between the muon transmittance and the density-length viewed by MT2 at $\theta = 67.2°$. We found the relation can be well approximated by this formula. Figure 19 plots the coefficient $a(\theta)$ as a function of the zenith angle.

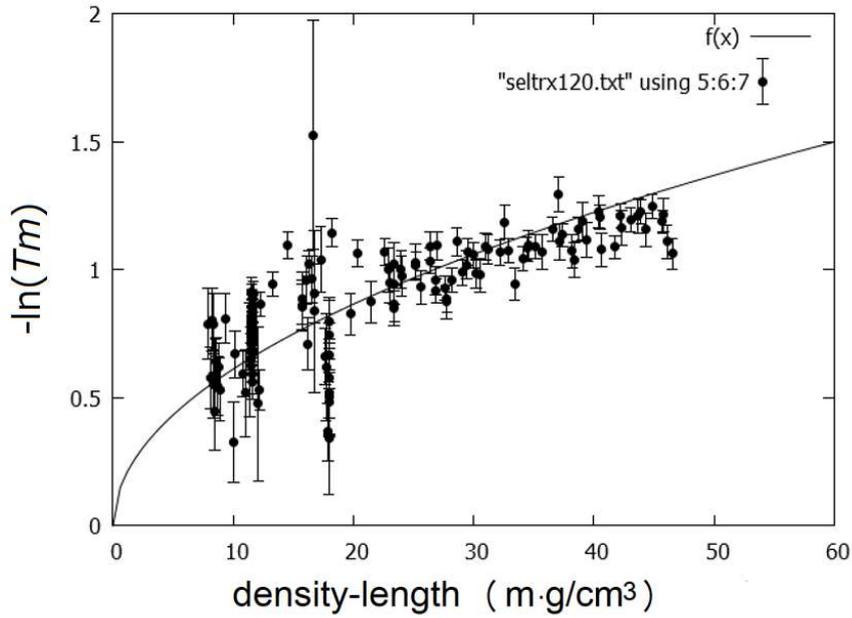

Fig. 18: Relationship between the calculated density-length and the measured muon transmittance at $\theta = 67.2°$. The curve is a fit to the formula (3).

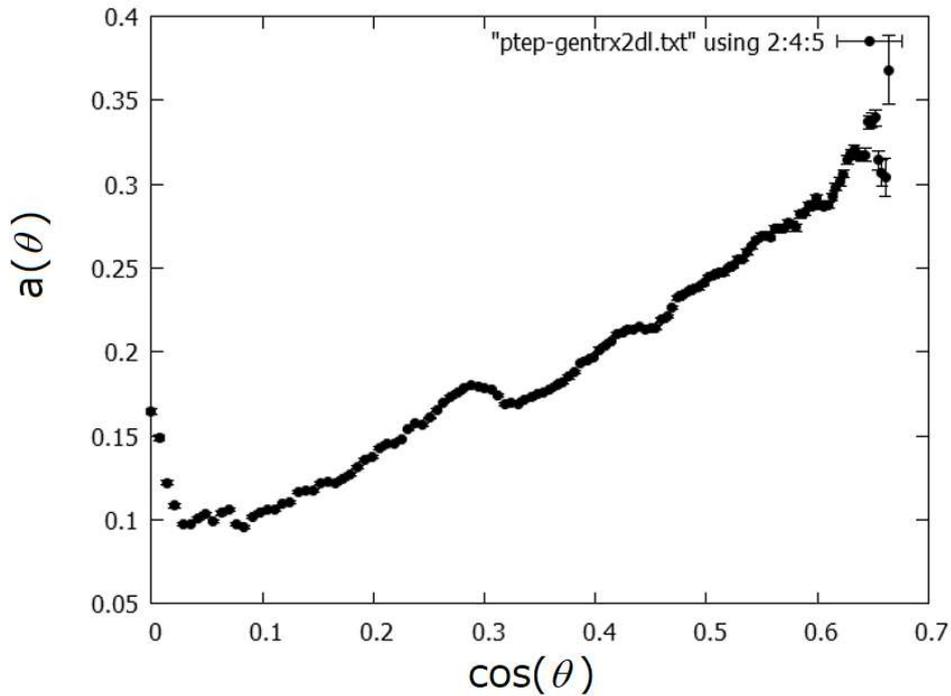

Fig. 19: The coefficient $a(\theta)$ obtained from the measured muon transmittance and the density-length distributions calculated for known objects in the region excluding the NFSP.

In Fig. 20 we show the density length distribution observed by the MT2, derived based on Eq. (3) using the measured muon transmittance and the coefficient $a(\theta)$. The upper plot shows the distribution in the entire view area. The lower plot is the density-length distribution inside the NFSP where the density-length of the objects outside the NFSP is subtracted from the top plot.

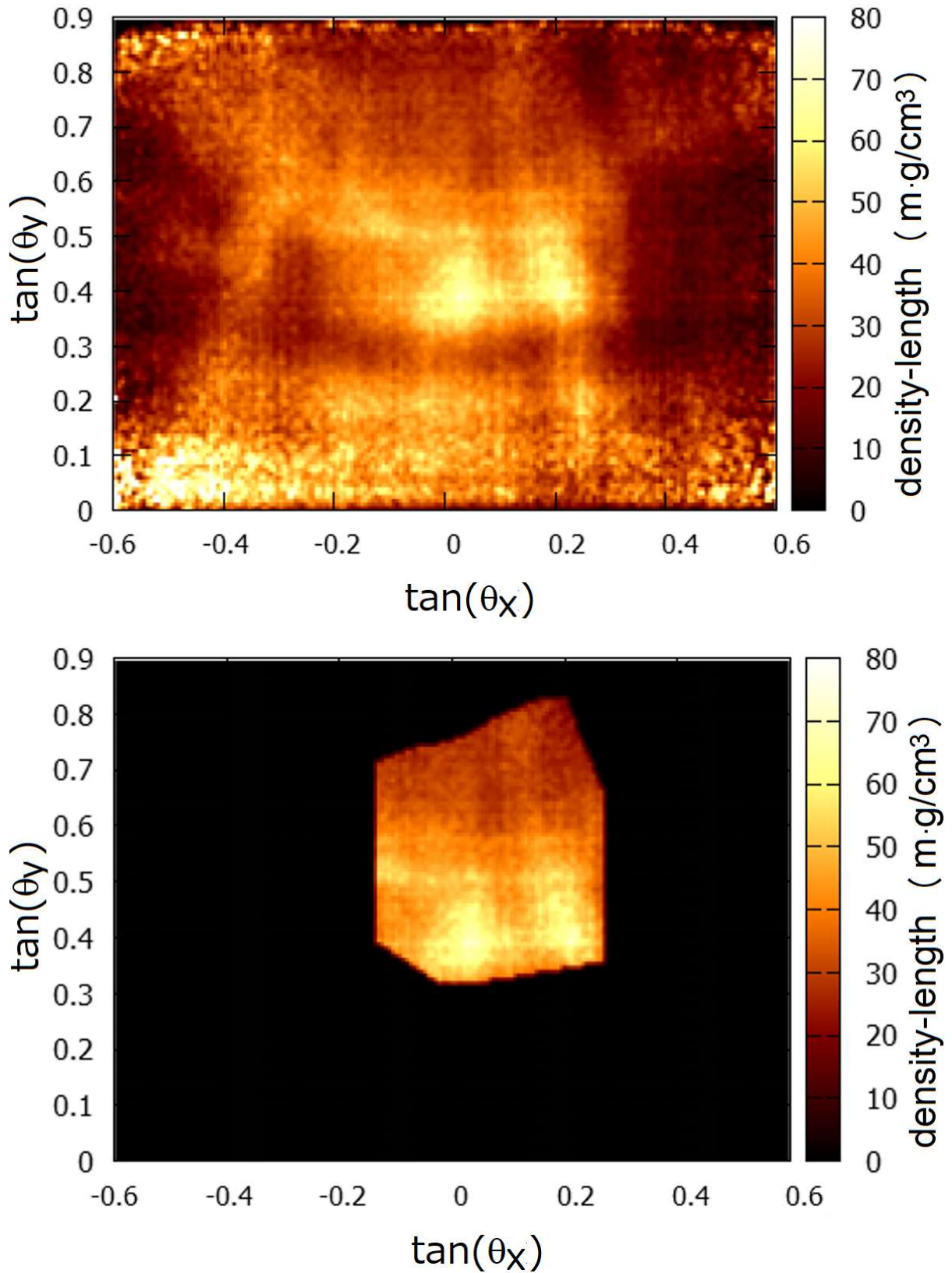

Fig 20. The density length distributions estimated from the MT2 data, (top) in the entire view area and (bottom) inside the NFSP.

The final procedure is the evaluation of the average density of the heavy object clusters shown in Fig. 16. We assume that both clusters are rectangular located in the NFSP as shown in Fig. 21.

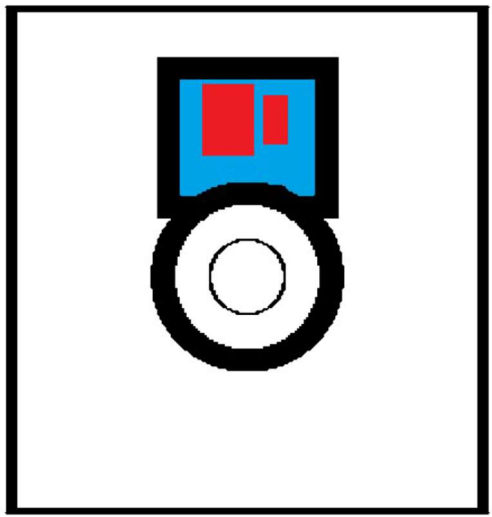

Fig 21: Assumed cross sectional view of the reactor at the height of the NFSP, showing the rectangular reactor building wall, the circular containment and pressure vessels, and the NFSP (water in blue). The locations and sizes of the observed heavy objects (red squares) are to be adjusted to reproduce the measured data.

Figure 22 shows the obtained density-length distribution inside the NFSP as a function of horizontal direction tan($\theta$x) in comparison with the estimates varying the average density of the heavy clusters at $\theta \sim 68°$. Note that $\theta$ is not constant as tan($\theta$y) is fixed in this plot, see Fig. 10. The comparison in the eitire NFSP region results the following sizes and density are optimum for the two heavy objects:

Cluster 1 (left rectangle in Fig. 21): 4.5 m (in left-right)×6.5 m (in top-bottom) and 4 m in height
Cluster 2 (right rectangle in Fig. 21): 3 m (in left-right)×5.4 m (in top-bottom) and 4 m in height
Average density (common for both): 3.5± 0.5 g/cm³.

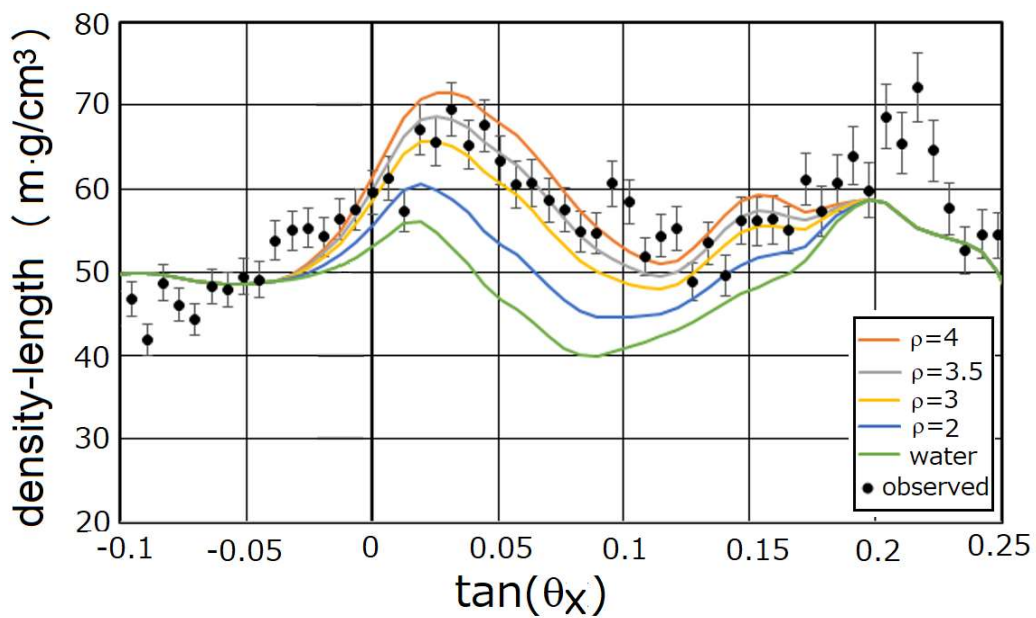

Fig 22: The density-length distribution inside the NFSP compared with estimates using various density, $\rho$, of the heavy object clusters.

Therefore the weights of the heavy clusters are:

Cluster 1: 410± 60 tons
Cluster 2: 227± 35 tons
In total: 637± 95 tons

For a typical water occupation of 65% in volume, the above weight refers to a net weight of 519 tons for the fuel assemblies alone and the density 2.8 g/cm$^3$ excluding water. This corresponds to 2000-2100 fuel assemblies located in the NFSP for a typical unit weight of 250 kg.

In the previous paper [3], we estimated using the MT1-1 data that the (un-separated) fuel cluster extended 6 m in top-bottom (in Fig. 21) and 7 m in left-right (in Fig. 21) under an assumption of the average density of 2.5 g/cm$^3$. Since the clusters viewed from MT-1 are not separated, the previous observation is consistent with the present estimates. Also the assumed density 2.5 g/cm$^3$ is consistent with 2.8 g/cm$^3$ by taking the uncertainty into account.

4. Conclusions

We investigated the inner structure of the nuclear reactor at the JAPC, Tokai, Ibaraki, Japan, with the muon radiography system by using the cosmic muons. The detectors were placed outside of the reactor building. We succeeded in observing the containment vessel, pressure vessel, and other components of the reactor building. From the three views, we also succeeded in locating two heavy object clusters which are considered to be the nuclear fuel assemblies stored in the nuclear fuel storage pool. The mass of the fuel assemblies was derived using the density-length measurement.

**Acknowledgements**

We thank the Japan Atomic Power Company (JAPC) for undertstanding the importance of this study, and for their cooperation and supports in carrying out this project. We thank Professor Hirotaka Sugawara of the Okinawa Institute of Science and Technology for his advice and suggestions.